# A novel method for calibration and monitoring of time synchronization of TOF-PET scanners by means of cosmic rays

Michał Silarski[1], Eryk Czerwiński[1], Tomasz Bednarski[1], Paweł Moskal[1], Piotr Białas[1], Łukasz Kapłon[1,2], Andrzej Kochanowski[2], Grzegorz Korcyl[1], Jakub Kowal[1], Paweł Kowalski[3], Tomasz Kozik[1], Wojciech Krzemień[1], Marcin Molenda[2], Szymon Niedźwiecki[1], Marek Pałka[1], Monika Pawlik[1], Lech Raczyński[3], Zbigniew Rudy[1], Piotr Salabura[1], Neha Gupta Sharma[1], Artur Słomski[1], Jerzy Smyrski[1], Adam Strzelecki[1], Wojciech Wiślicki[3], Marcin Zieliński[1] and Natalia Zoń[1]

[1] Institute of Physics, Jagiellonian University, Kraków, Poland

[2] Faculty of Chemistry, Jagiellonian University, Kraków, Poland

[3] Świerk Computing Centre, National Centre for Nuclear Research, Otwock-Świerk, Poland

Corresponding author: Michal Silarski, Institute of Physics, Jagiellonian University, 30-059 Kraków, Poland

E-mail: michal.silarski@uj.edu.pl

**Abstract:** All of the present methods for calibration and monitoring of TOF-PET scanner detectors utilize radioactive isotopes such as e.g. $^{22}$Na or $^{68}$Ge, which are placed or rotate inside the scanner. In this article we describe a novel method based on the cosmic rays applica-tion to the PET calibration and monitoring methods. The concept allows to overcome many of the drawbacks of the present methods and it is well suited for newly developed TOF-PET scanners with a large longitudinal field of view. The method enables also monitoring of the quality of the scintillator materials and in general allows for the continuous quality assurance of the PET detector performance.

## Introduction

Positron emission tomography (PET) is one of the most dynamically developing diagnostic methods that allows for non-invasive imaging of physiological processes occurring in the body. It permits to determine the spatial and temporal distribution of concentrations of selected substances in the body. Typical PET detectors are built out of hundreds of detection modules (see e.g. Figure 1) which enable registration of the annihilation gamma quanta in coincidence. In the newer generation of PET detectors the resolution of the tomographic image is improved by determination of the annihilation point along the line-of-response (LOR) [1-2]. It is based on measurements of the time difference between the arrivals of the gamma quanta to the detectors. This technique is known as TOF (time of flight), and improves the reconstruction of PET images by increasing signal to noise ratio due to the reduction of noise propagation along the LOR during the reconstruction [3-4].

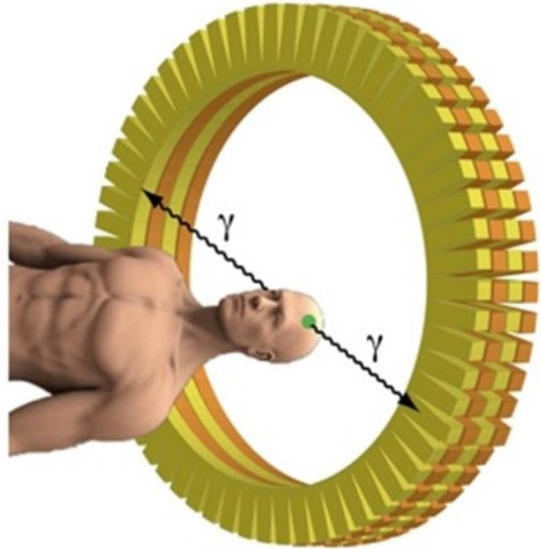

Fig. 1 Schematic illustration of the standard PET detector system.

Figure adapted from [5].

Reconstruction of the tomographic image would not be possible if all the PET detectors were not be synchronized in time and calibrated in view of energy measurements. There are many methods for determination of the time and energy calibration constants and monitoring of the PET scanner detectors. Currently, calibrations are performed using radioactive isotopes such as $^{22}$Na or $^{68}$Ge, which are placed inside the PET scanner in precisely defined locations, for example in the geometric center. The radioactive source can be covered with metal or plastic shield causing gamma quanta scattering and, as a consequence, enabling synchronization of all PET detectors [6, 7]. There are methods for time synchronization of TOF-PET detectors which uses several radioactive sources simultaneously, enabling calibration even during scanning of the patient. Gamma quanta originating from radioactive sources are identified based on the known positions and the time information from the detectors, which allows for the rejection of these events in the tomographic image reconstruction [8]. One can use also radioactive source rotating along the scintillation chamber, which allows relative synchronization using the fact that the time difference registered by opposite detectors is constant [9]. However, the methods described above do not allow for simultaneous calibration of detectors during the scanning process without exposing a patient to an additional radiation dose emitted by the radioactive sources. Furthermore, the use of radioactive sources for TOF-PET detectors synchronization requires additional equipment, trained personnel and the replacement of resources (for example, the half-life of $^{68}$Ge is approximately 270 days), which increases the cost of imaging. Currently, calibration is generally performed once per day, before scans of patients, to prevent exposure to the additional radiation dose. This however makes impossible to take into account environmental conditions such as e.g. temperature fluctuations that affect the time and energy properties of detectors. Moreover, usage of current methods for calibrating TOF-PET detectors is not convenient for long detectors used in the new STRIP- and MATRIX-PET's [10-11], in which the polymer material in form of long scintillator bars or plates is used (see Figure 2). Taking into account drawbacks of the described methods, especially in view of the novel scanner concepts, development of new calibration techniques is needed. As it is presented in next sections, to this end one may use cosmic rays constituting an inexhaustible source of radioactivity.

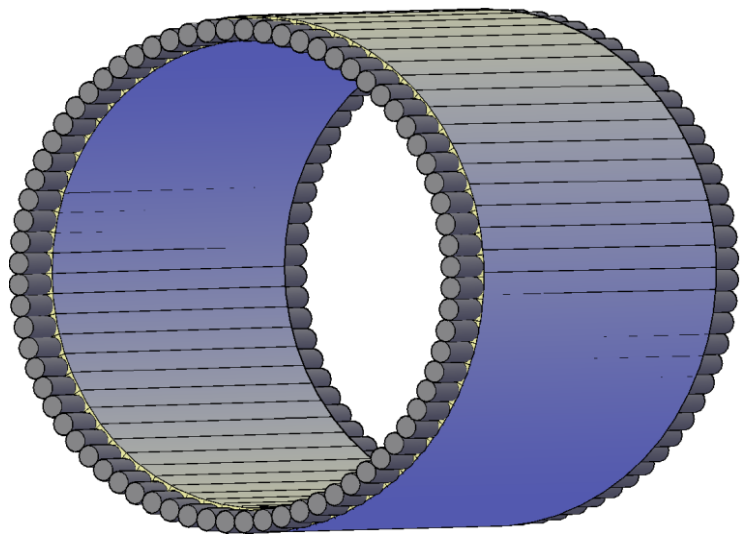

Fig. 2 Schematic view of the STRIP-PET diagnostic chamber consisting of the plastic scintillator strips read out on two sides by photomultipliers.

# Cosmic rays as a tool for the PET scanners calibration

The atmosphere of earth is bombarded with elementary particles and light nuclei originating mainly from solar flares, as well as from outside the solar system. A small fraction of primary cosmic protons and electrons with high energy reach the surface, but the majority of cosmic rays are absorbed in the atmosphere leading to production of secondary particles. The most numerous particles at the sea level are muons with integral vertical intensity of $I \approx 70$ m$^{-2}$s$^{-1}$sr$^{-1}$ [12].

In the presently used PET scanners cosmic rays have been considered as a source of background, relatively easy to reject due to much bigger energy deposits in the scintillating crystals with respect to the annihilation quanta. However, since the energy and angular distributions for cosmic rays are known with a good precision, this background may be useful in the synchronization and energy calibration procedures for tomographs, especially for STRIP-PET with a large longitudinal field of view.

## Time synchronization of PET detection modules

In STRIP-PET scanner, lines-of-response are determined based on the reconstructed position $x$ along the scintillator strip (Figure 3). Therefore, the time synchronization has to be

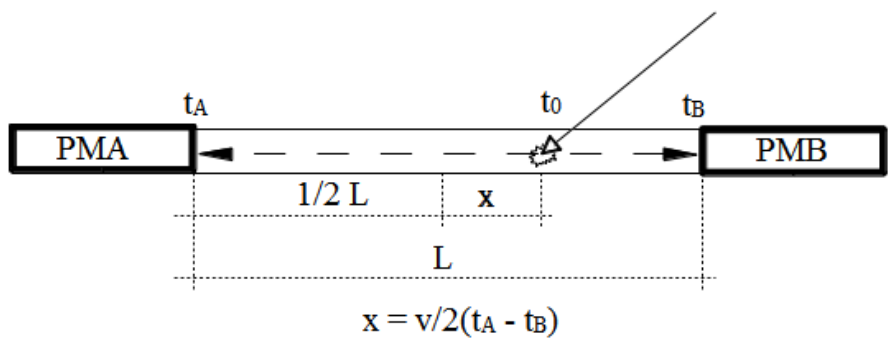

Figure 3 Schematic view of a single detector module used in the STRIP-PET.

Position where the gamma quantum interacted can be determined from the difference between times measured at both edges of the strip. L denotes the length of the scintillator strip, and x denotes the distance along the strip between the center of the scintillator and the interaction point.

done first for every detection module separately. The times registered on sides A and B of a module can be expressed as: $t_{A(B)} = t_0 + t^p_{A(B)} + t^o_{A(B)}$, where $t_0$ is the true interaction time of gamma quantum (or e.g. cosmic muon), $t^p_{A(B)}$ denotes the light propagation time to side A(B) of the detector and $t^o_{A(B)}$ is the time offset due to e.g. front-end-electronics or signal formation in the photomultiplier PMA(B). The hit position $x$ can be determined from the difference between times measured at both edges of the strip $\Delta t_{AB}$:

$$x = \frac{v}{2}(\Delta t_{AB} - \Delta t^0_{AB}), \quad (1)$$

where $v$ denotes the effecitive velocity of light in the scintillator. To determine both, $v$ and the relative offset between measured times $\Delta t^0_{AB}$ one can utilize the fact, that every scintillating detector of the scanner barrel (Figure 2) is uniformly exposed, in terms of $x$, for cosmic rays. Thus, the time diference distribution $\Delta t_{AB}$ should be uniform and confined in the following range (assuming infinite time measurement precision): $\Delta t_{AB} \in \left[\Delta t^o_{AB} - \frac{L}{v}; \Delta t^o_{AB} + \frac{L}{v}\right]$, where $L$ denotes the length of the scintillator. This spectrum can be fitted with a theoretical function, for example with a product of two Fermi functions, giving estimates of the velocity $v$ and offset $\Delta t^o_{AB}$. This allows for synchronization of channels within a single detector module and, as a consequence, allows to determine the reaction point along the scintillator strip.

Synchronization of different scanner detectors can be made by measuring the time of flight of cosmic rays for all pairs of scintillator strips separated by a minimum distance, for example by at least 30 cm. Knowing the reaction point for each detector one can determine the distance $d$ traveled by the muon between the two modules. Since at the Earth's surface the cosmic rays velocity distribution is well known, one can estimate the average time of flight of the muon for each pair of modules. The difference between the measured time of flight of the cosmic rays and the expected value gives an estimate of the relative delay between the two PET scanner modules (labeled as 1 and 2):

$$\Delta t_{12}^o = t_1 - t_2 - t_{12}^{TOF} = \frac{1}{2}(t_{1A} + t_{1B}) - \frac{1}{2}(t_{2A} + t_{2B}) - t_{12}^{TOF}, \quad (2)$$

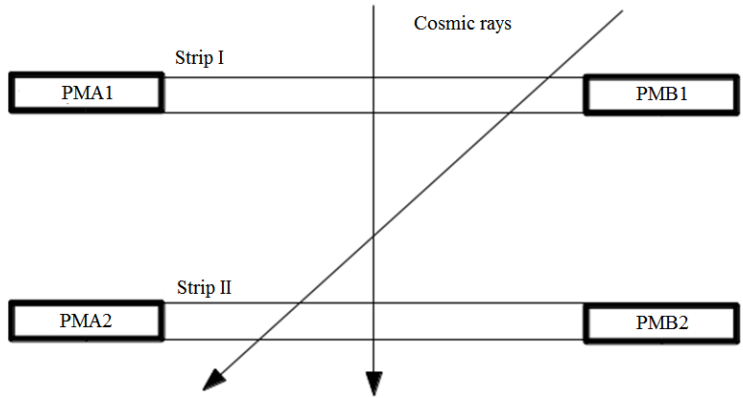

Fig. 4 Visualization of two cosmic ray particles passing at different angles through PET scintillation detector modules.

where $t_{12}^{TOF}$ is the average time of flight of cosmic rays. In order to obtain high precision of the calibration the procedure should be performed for cosmic rays passing through the modules at different angles (see Figure 4) and for all pairs of modules providing set of parameters to ensure the global synchronization of the scanner.

Synchronization can be accomplished also without the use of information about the velocity distribution of muons reaching the Earth's surface. To this end for every pair of detectors one has to determine the reference spectra of cosmic rays time-of-flight or alternatively of veloci-

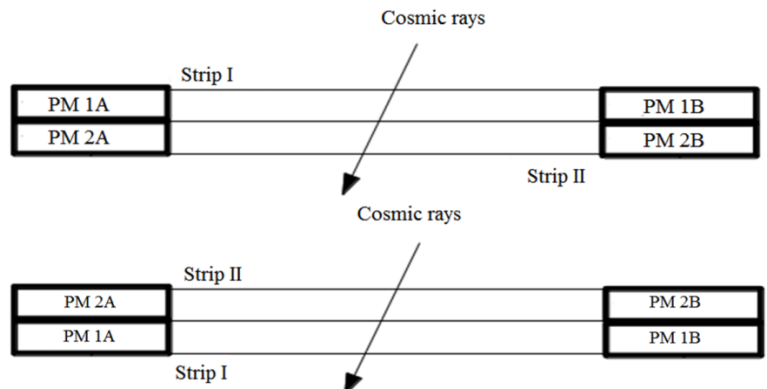

Fig. 5 Method of determination of the relative time offset between two detection modules.

First, one measure the spectrum of time difference between strips 1 and 2 (upper picture) and then, after the change of detector order, between strips 2 and 1 (lower picture).
ty. In this case the synchronization of the scanner is done by setting the relative time delays between pairs of modules so that the measured spectra of muons velocity or time of flight for each pair of strips fit the reference spectra. Here the relative delays are free parameters of the adjustment. One way to determine the reference spectra is shown in Figure 5. The two strips are placed first one on top of the other, then a measurement of spectra of the time difference between the upper and lower strip is performed. Next the order of detectors is reversed and one repeats the measurement. This allows for determination of the relative time offset between two detection modules. The two detectors are next placed at distance and horizontal angle corresponding to their position in the scanner, and the measurement of the corresponding spectra is repeated.

**Energy calibration of PET detection modules**

The use of cosmic rays allows also for energy calibration of PET scanner. The light signals reaching the two ends of a scintillating strip are converted by photomultipliers to electric pulses with charge $Q_{A(B)}$. The value of $Q_{A(B)}$ depends on the energy $E_d$ deposited in the scintillator and the hit position $x$ in the strip:

$$Q_{A(B)}(E_d, x) = \beta_{A(B)} f(x) E_d. \quad (3)$$

The aim of the calibration is the determination of constants $\beta_{A(B)}$ expressing the quantum efficiency of photomultipliers photocathodes and gains, as well as determination of the function $f(x)$. Therefore, energetic calibration should be divided into two separate procedures: (i) monitoring and determining of the new value of high voltage supplied to photomultipliers to keep all gains for the whole scanner approximately equal, and (ii) determining values of constants allowing to calculate the absolute values of energy deposited in scintillators on the basis of measured values of the charge of registered signals. The form of analytical approximation of $f(x)$ does not depend on the detection module and it is usually expressed as $f(x) =$

$e^{-\lambda\left(x+\frac{L}{2}\right)}$, where $\lambda$ is the effective light attenuation length of scintillator. This approximation is very good except for a few centimeter distance near photomultipliers. However, in practice this function can be determined directly for each detection module, for example by measuring the charge $Q_{A(B)}$ for the same energy deposited and for different well-defined positions $x$. The standardization procedure may involve the comparison of charge spectra of signals for which the hit position of cosmic rays is close to the center of the detector strip with reference spectra measured for every angular arrangement separately. Then, based on the gain dependence on the applied voltage known for each photomultiplier, one calculates new power supply voltage. For quick monitoring of gains for a single strip one can take advantage of the following relation between the ratio of registered charges and position along the scintillator $x$:

$$ln\left(\frac{Q_A}{Q_B}\right)(x) = -2\lambda x + ln\left(\frac{\beta_A}{\beta_B}\right). \quad (4)$$

The above formula was derived based on equation (3). Fitting a linear function to the measured values of $ln\left(\frac{Q_A}{Q_B}\right)$, excluding the scintillator regions in vicinity of photomultipliers, allows for the direct estimation of the effective length of the light absorption $\lambda$. Thus, it allows for monitoring of the quality of the detector material. Moreover, if the gains of photomultipliers are equal, the following relation holds: $ln\left(\frac{\beta_A}{\beta_B}\right) = 0$. This gives possibility to monitor and correct calibration of gains of photomultipliers of a single module.

Alternatively, one can determine the $ln\left(\frac{Q_A}{Q_B}\right)$ distribution omitting the dependence on $x$, as in the case of time calibration taking advantage that each strip is uniformly exposed by cosmic rays. This distribution should be also uniform and can be fitted with a theoretical function e.g. of the form:

$$f\left(\ln\left(\frac{Q_A}{Q_B}\right)\right)$$

$$= \frac{N}{\left\{1 + exp\left[\left(-\ln\left(\frac{Q_A}{Q_B}\right) + \lambda L + \ln\left(\frac{\beta_A}{\beta_B}\right)\right)/\sigma_Q\right]\right\}\left\{1 + exp\left[\left(\ln\left(\frac{Q_A}{Q_B}\right) - \lambda L - \ln\left(\frac{\beta_A}{\beta_B}\right)\right)/\sigma_Q\right]\right\}}, \quad (5)$$

where $N$ is an normalization factor and $\sigma_Q$ express the precision of charge measurement. Equality of gains requires that this distribution is symmetric with respect to $\ln\left(\frac{Q_A}{Q_B}\right) = 0$, which implies that for the equalization of photomultiplier gains we can again monitor the value of $\ln\left(\frac{\beta_A}{\beta_B}\right)$.

The second step of the calibration procedure consists of determining calibration constants allowing to convert charges of measured signals into the absolute values of energy deposited in the scintillator. The absolute energy scale can be determined from the geometrical average of charge measured on both sides of the strip $\sqrt{Q_A Q_B}$, which is proportional to the energy $E_d$ deposited in the scintillator: $\sqrt{Q_A Q_B} = \beta E_d e^{(-\lambda L/2)}$. Thus, $\sqrt{Q_A Q_B} C = E_d$, where $C$ is a calibration constant to be determined for each detector module.

Since the energy deposited in the strip depends on the length of the cosmic ray path inside the scintillator the path length should be controlled e.g. by determining distribution of $\sqrt{Q_A Q_B}$ for cosmic rays passing the detector vertically in a small solid angle θ around 0°. One can also take a wider range of well-defined angles and normalize the measured $\sqrt{Q_A Q_B}$ distribution to the path length of the radiation passing the scintillator. The energy deposits $E_d$ normalized to the path length of the cosmic rays are well known for many materials, in particular for plastic scintillators [10]. Since effectively the distribution of energy losses of cosmic ray muons depends on the θ angle, one has to discretize the range of θ and estimate an average $\langle\sqrt{Q_A Q_B}\rangle$ and $\langle E_d\rangle$ for every angular bin. The calibration constant could be then calculated using the following formula: $\langle\sqrt{Q_A Q_B}\rangle C = \langle E_d \rangle$. As the final estimate of $C$ one can take a weighted average of values determined for different angular ranges. Finding values of $\langle\sqrt{Q_A Q_B}\rangle$ and $\langle E_d \rangle$ for several θ ranges we can monitor the systematic uncertainty of determining the calibration constant $C$, which has to be determined for each detector module independently. Absolute calibration of the energy, as in the case of time synchronization, may be also performed without knowledge about the energy loss distribution of muons. In this case, before the construction of the PET scanner every detector module should be calibrated using several different radioactive sources. One can then measure the reference spectra of the energy of the cosmic rays deposited in scintillation strips. Later, in an already built and operating tomograph, the calibration would consist in determining the calibration constant $C$ so that the spectra collected for the cosmic rays during the calibration agree with the reference spectrum measured before.

**Experimental results for a single detection module**

In the Institute of Physics of the Jagiellonian University, in the framework of the J-PET Collaboration, we test calibration methods described in the last section with a single module of a prototype of STRIP-PET. The module was built of 30 cm BC-420 scintillator strip with 19 x 5

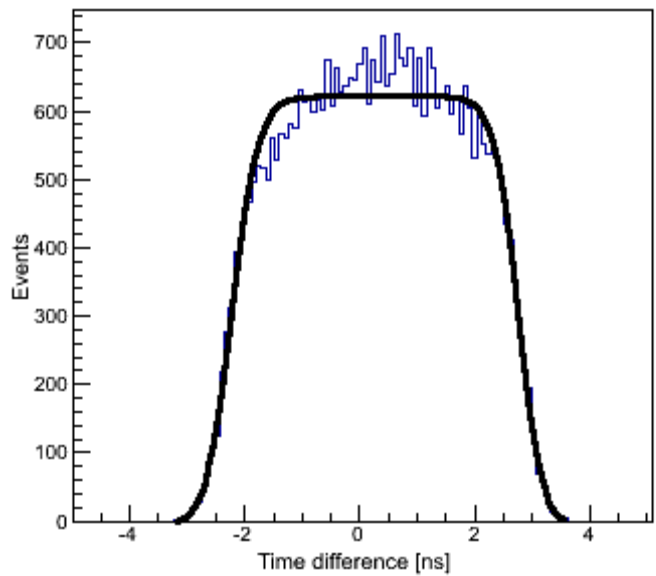

Fig. 6 Distribution of the time difference of signals registered at both ends of the scintillator strip. The fitted curve is described in the text.

mm$^2$ cross section wrapped with tyvek foil. The scintillator was read out by two R5320 Hamamatsu photomultipliers with equalized gains. We have performed a six-hour cosmic ray run acquiring about 40000 events. Using the LeCroy WaveRunner 64Xi oscilloscope we have recorded full signals for both photomultipliers sampled with 50 ps intervals, triggering events only if signals in both photomultipliers were in coincidence. The times of signals were then determined for all signals at the constant level of -0.6 V. The distribution of difference of times measured on both ends of the module is presented in Figure 6, where one observe a shift towards bigger values of time differences. This distribution was fitted with a function of the following form:

$$N(\Delta t_{AB}) = \frac{A}{\left\{1 + exp\left[\left(-\Delta t_{AB} + \Delta t_{AB}^o - \frac{L}{v}\right)/\sigma_{t1}\right]\right\}\left\{1 + exp\left[\left(\Delta t_{AB} - \Delta_{AB}^o - \frac{L}{v}\right)/\sigma_{t2}\right]\right\}}, \quad (6)$$

where $A$ is an normalization factor and $\Delta t_{AB}^o$ denotes the relative time offset. $\sigma_{t1}$ and $\sigma_{t2}$ parameters express the precision of $\Delta t_{AB}$ measurement. These two parameters should be in principle equal, but in general, due to e.g. scintillator heterogeneities, their values may be different. The results of the fit are gathered in Table 1.

| Parameter of the fit | Value |
|---|---|
| $A$ | 157.7 ± 1.4 |
| $\Delta t_{AB}^o - L/v$ | -2.216 ± 0.012 [ns] |
| $-\Delta_{AB}^o - L/v$ | -2.755 ± 0.001 [ns] |
| $\sigma_{t1}$ | 0.234 ± 0.009 [ns] |
| $\sigma_{t2}$ | 0.205 ± 0.007 [ns] |

Tab. 1 Values of parameters obtained with the fit to the measured $\Delta t_{AB}^o$ distribution.

They allow to determine both the time offset $\Delta t_{AB}^o$ =0.270 ± 0.02 [ns], and the light propagation velocity in the strip $v$ =12.1 ± 0.2 [cm/ns]. It has to be stressed, that the estimated values of parameters are consistent with results obtained with other methods used in studies of one module prototype of STRIP-PET. Moreover, with the same setup we have made measurements using collimated beam of annihilation gamma quanta originating from $^{68}$Ge radioactive source. In Figure 7 we compare the charge spectra (in terms of registered number of photoelectrons) obtained with the $^{68}$Ge source and cosmic rays for one of the photomultipliers used in the measurements. The scintillator strip was irradiated with the beam of gamma quanta collimated (with about 3mm spread) on the center of the strip. In case of cosmic muons only events with -0.25 ns $\leq \Delta t_{AB} \leq$ 0.25 ns were taken into account. This corresponds to the region of about 3 cm around the center of the strip. The maximum for number of photoelectrons around 250 corresponds to the mean energy deposition of muons at the surface in the polyvinyltoluene, equal to about 2.32 MeV/cm. One can see that, in principle, the signals from annihilation quanta could be distinguished from cosmic rays interactions based only on the charge of signal from one photomultiplier. For a clear separation one should, however, use some other methods, for example requiring for cosmic rays an appropriately large signals in two modules (four photomultipliers) of PET scanner.

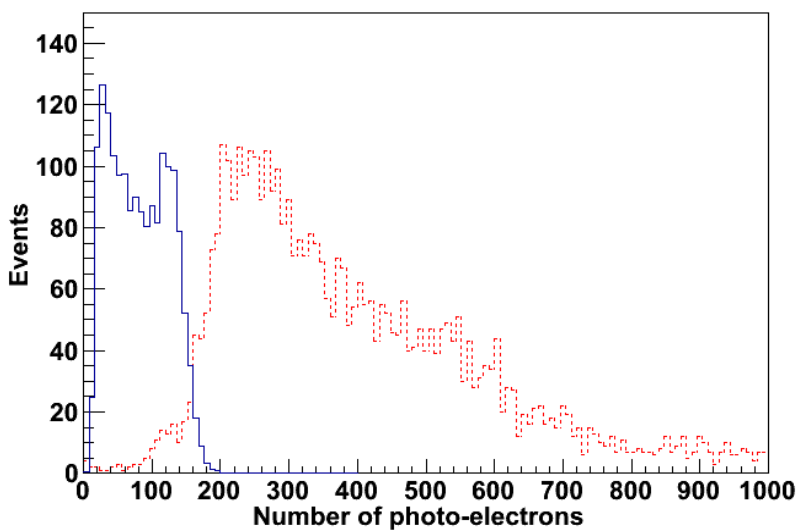

Fig. 7 Comparison between charge spectra registered using annihilation quanta from $^{68}$Ge source (solid histogram) and cosmic rays (dotted histogram). Further description can be found in the text.

## Conclusions

Cosmic rays provide an inexhaustible source of radiation which may be used for synchronization and calibration of TOF-PET scanner detectors. Monitoring of the scanner can be performed during and between examinations without exposing patients to the additional radiation doses. Signals from cosmic muons can be easily distinguished from annihilation gamma quanta, since they deposit much more energy in the scintillator material. The calibration constants may be monitored every couple of hours using the method described in this article. Moreover, the whole calibration procedure could be performed automatically without involving any specialized staff and any additional costs of radioactive sources. Presented method is especially suited for calibration of the novel STRIP-PET scanners [9-11] which are being developed in the Institute of Physics of the Jagiellonian University, and for other PET detectors with large longitudinal field of view e.g. [13-16].

## Acknowledgments

We acknowledge technical and administrative support by M. Adamczyk, T. Gucwa-Ryś, A. Heczko, M. Kajetanowicz, G. Konopka-Cupiał, J. Majewski, W. Migdał, A. Misiak, and the financial support by the Polish National Center for Development and Research through grant INNOTECH-K1/IN1/64/159174/NCBR/12, the Foundation for Polish Science through MPD programme and the EU and MSHE Grant No. POIG.02.03.00-161 00-013/09.

**Conflict of interest statement**

**Authors' conflict of interest disclosure:** The authors stated that there are no conflicts of interest regarding the publication of this article.

**Research funding:** None declared.

**Employment or leadership:** None declared.

**Honorarium:** None declared.